\documentstyle[11pt,amsfonts]{article}
\renewcommand{\theequation}{\arabic{equation}}
\newcommand{\be}{\begin{equation}}
\newcommand{\ee}{\end{equation}}
\newcommand{\bea}{\begin{array}}
\newcommand{\ea}{\end{array}}
\newcommand{\beqa}{\begin{eqnarray}}
\newcommand{\eeqa}{\end{eqnarray}}
\newcommand{\bean}{\begin{eqnarray*}}
\newcommand{\eean}{\end{eqnarray*}}

\def\up#1{\leavevmode \raise.16ex\hbox{#1}}

\newcommand{\gapproxeq}{\lower
 .7ex\hbox{$\;\stackrel{\textstyle >}{\sim}\;$}}
\newcommand{\lapproxeq}{\lower .7ex\hbox{$\;\stackrel
{\textstyle <}{\sim}\;$}}

\renewcommand{\theequation}{\thesection.\arabic{equation}}

\newcounter{appendice}
\newcommand{\appendice}
{
\setcounter{equation}{0}
\renewcommand{\theequation}{\Alph{appendice}.\arabic{equation}}
\addtocounter{appendice}{1}
{\Large{\bf  Appendix \Alph{appendice}}}
}

\def\thebibliography#1{{\bf REFERENCES\markboth
 {REFERENCES}{REFERENCES}}\list
 {[\arabic{enumi}]}{\settowidth\labelwidth{[#1]}\leftmargin\labelwidth
 \advance\leftmargin\labelsep
 \usecounter{enumi}}
 \def\newblock{\hskip .11em plus .33em minus -.07em}
 \sloppy
 \sfcode`\.=1000\relax}

\def\BI{{\rm 1\!l}}

\begin{document}
\vskip 1cm
\centerline{ \LARGE Snyder space revisited }

\vskip 2cm

\centerline{{\sc Lei Lu}\footnote{llv1@crimson.ua.edu }  and    {\sc A. Stern }\footnote{astern@bama.ua.edu}   }

\vskip 5mm

\centerline{  Department of Physics and Astronomy, University of Alabama,
Tuscaloosa, AL 35487, U.S.A.}
\vskip 2cm

\vspace*{5mm}

\normalsize
\centerline{\bf ABSTRACT}

We examine basis functions on momentum space  for the three-dimensional Euclidean Snyder algebra.   We argue that the  momentum space is isomorphic to the $SO(3)$ group manifold, and  that the basis functions span either one of two  Hilbert spaces.  This  implies the existence of two distinct lattice structures of space. Continuous rotations and translations are unitarily implementable on these lattices.

\vskip 4 cm 
\newpage

\section{Introduction}

 In 1947, Snyder wrote down a Lorentz covariant deformation of the Heisenberg algebra, with the properties that the position operators are noncommuting and have  discrete spectra.\cite{Snyder:1946qz}  
From the discrete position spectra, representations of the algebra imply a lattice description of space, which we refer to as  `Snyder space'.   The lattice here differs from that used to write down lattice field theories, because  only one coordinate  can be determined in a measurement, and the covariance indicates that the lattice is compatible with the continuous symmetry transformations of space-time.
 
 In this article we examine representation theory for the Snyder algebra,  along with the implementation of continuous transformations on the lattice.
 Because the time component of the space-time four vector in Snyder's algebra has a  continuous spectrum, we shall focus on  
  the subalgebra generated by three position and momentum  operators.  Angular momentum operators  are constructed from the position and momentum  operators in the usual way, and they   generate the $SO(3)$ rotation group.
  The group is enlarged to   
 $SO(4)$ upon including the position operators along with the  angular momentum.  The discrete spectra for the position operators easily follows from the discreteness of the $SO(4)$ representations.  
The corresponding $so(4)$  algebra contains only one independent quadratic Casimir operator, and  so  unitary irreducible representations of $SO(4)$ that occur for this model are labeled by  a single  quantum number $j$, which a priori can have integer or half-integer values.
The  $SO(4)$ representations that result from the full Snyder algebra are  infinite-dimensional reducible representations.   This is due to the action of the group in momentum space.   Group representations,  in general,  depend on the topology of the underlying space.  In a related problem, the topology of momentum space played an important role in obtaining the position spectra  for point particles in $2+1$ gravity.\cite{'tHooft:1996uc},\cite{Matschull:1997du} For the case of the three-dimensional Euclidean Snyder algebra, we argue that one gets a consistent  action of $SO(4)$ upon making an  identification at infinite momentum whereby momentum space is isomorphic to the $SO(3)$ group manifold.  
Because $SO(3)$ is doubly connected, the quantum theory is not unique.  We find two sets of basis functions on momentum space, spanning two distinct Hilbert spaces ${\cal H}_B$  and ${\cal H}_F$.  The  basis functions are distinguished  by 
 their asymptotic properties and also by the  quantum number   $j$, which takes all integer values for ${\cal H}_B$, and all half-integer values for ${\cal H}_F$.  Because the values of $j$ determine the spectra of the position operators, the two Hilbert spaces imply the existence of  two different spatial lattices, i.e.,  two different Snyder spaces.  Rotations and translations are implemented as unitary transformations on the lattices.  In fact, the full Poincar\'e group (and possibly even the super-Poincar\'e group) can be made to have a consistent action on the lattices.  This will be shown in forthcoming works by considering relativistic particle dynamics on Snyder space.\cite{leilu}  Nonrelativistic dynamics on three-dimensional Snyder space was recently considered  in \cite{Mignemi:2011gr}.

The outline for the article is the following.  In section 2, we review the three-dimensional (Euclidean) Snyder  algebra and obtain the spectra for the  position operators.   We consider position operators associated with both Cartesian, and spherical coordinates (more specifically, we find the eigenvalues of the radial coordinate).   The action of the $SO(4)$ group on momentum space is discussed in section 3.
We first write down a one parameter family of differential representations for the position operators,  and then impose asymptotic conditions in momentum space.
We obtain basis functions associated with the Hilbert spaces ${\cal H}_B$  and ${\cal H}_F$ in section 4.
Eigenfunctions  of  the radial coordinate form a convenient set of such basis functions and they correspond to  spherical harmonics  on $S_+^3$, meaning $S^3$  restricted to one hemisphere.  We also discuss momentum  eigenfunctions of Cartesian coordinate operators.    In section 5 we write down unitary transformations on the lattices associated with the  translation and rotation group.
Concluding remarks are made in section 6, including some speculations on the noncommutative field theory. 
In Appendix A we construct  normalizable wavepackets in momentum space which can be explicitly transformed to the discrete position space. 
Finally, there are several analogies between  Snyder space and the hydrogen atom, which we mention in Appendix B.  In both cases, $SO(4)$ is the relevant symmetry group,  there is one independent quadratic Casimir operator and  the $SO(4)$ representations that appear  are infinite-dimensional and  reducible.
  On the other hand,  $SO(4)$ acts differently on momentum space for the two systems.\cite{Bander:1965rz}   Unlike the case with Snyder's algebra,  momentum space for the hydrogen is topologically $S^3$. The momentum-dependent eigenfunctions associated with hydrogen atom bound states were found very long ago  by  Podolsky and Pauling\cite{Pauling} and by Fock\cite{Fock:1935vv}, and they are spherical harmonics on  $S^3$.

\section{Snyder Algebra and Position Eigenstates}

\setcounter{equation}{0}

\subsection{Three-dimensional Snyder  algebra}

The three-dimensional (Euclidean) Snyder  algebra is generated by position and momentum  operators,  $\hat x_i$ and  $\hat p_i$, $i=1,2,3$, respectively, which satisfies  commutation relations
\beqa [\hat x_i,\hat x_j] &=&\frac i{\Lambda^2}\epsilon_{ijk}\hat L_k\cr& & \cr [\hat x_i,\hat p_j] &=&i\biggl(\delta_{ij} +\frac{\hat p_i\hat p_j}{\Lambda^2}\biggr)\cr & &\cr [\hat p_i,\hat p_j] &=&0\label{snydrsubalg}\eeqa 
It is a deformation of the  Heisenberg algebra, where  $\Lambda\ne 0$ is the  deformation parameter with units of energy.
The Heisenberg algebra is recovered when $\Lambda\rightarrow \infty$.   $\hat L_i=\epsilon_{ijk}\hat x_j \hat p_k$ are the angular momenta.  They satisfy the  usual  commutation relations
\beqa [\hat x_i,\hat L_j] &=& i\epsilon_{ijk}\hat x_k\cr & & \cr [\hat p_i,\hat L_j] &=&i\epsilon_{ijk}\hat p_k\cr& &\cr [\hat L_i,\hat L_j] &=&i\epsilon_{ijk}\hat L_k\label{angmomalg}\;,\eeqa 
and so generate the three-dimensional rotation group.   In the following subsection, we  enlarge to the group to  
 $SO(4)$ by including $\hat x_i$  in the set of rotation generators.

Here and throughout the article we  assume that $\Lambda^2>0$.  The case of $\Lambda^2<0$ is relevant for double special relativity\cite{AmelinoCamelia:2010pd}
 and is characterized by an upper bound on the momentum and deformed energy-momentum dispersion relations.  In the latter case, $\hat x_i$ and  $\hat L_i$ generate an $SO(3,1)$ algebra and $\hat x_i$ have continuous eigenvalues.  For a discussion of nonrelativistic quantum mechanics for the case $\Lambda^2<0$, see \cite{Mignemi:2011gr}.

\subsection{so(4) algebra}

Here we examine the so(4) algebra generated by the  coordinates and angular momenta, and give the spectra of the position operators. We follow the discussion in \cite{Stern:2010ri}.

We define
 $\hat L_{AB}=-\hat L_{BA}$, with 
\be \hat L_{ij}=\epsilon_{ijk}\hat L_k \quad\qquad 
\hat L_{i4}=\Lambda \hat x_i \label{dflab}\ee
  Then (\ref{snydrsubalg}) and (\ref{angmomalg}) gives the standard form of the $so(4)$ algebra 
 \beqa 
[\hat L_{AB},\hat L_{CD}]&=& i(\delta_{AC} \hat L_{BD}-\delta_{BC} \hat L_{AD} -\delta_{AD} \hat L_{BC}+\delta_{BD} \hat L_{AC})\label{so(4)algbr}
\eeqa
Alternatively, we have the two $SU(2)$ generators
\be \hat A_i =\frac 12(\hat L_i +\Lambda \hat x_i)\qquad\quad \hat B_i =\frac 12(\hat L_i -\Lambda \hat x_i)\;,\ee  satisfying
\beqa [\hat A_i,\hat A_j] &=&i\epsilon_{ijk} \hat A_k\cr & &\cr  [\hat B_i, \hat B_j] &=&i\epsilon_{ijk} \hat B_k\cr & &\cr [\hat A_i,\hat B_j] &=&0 \; \label{ofralgbr} \eeqa 
From  (\ref {angmomalg}) one   has the identity
\be \hat x_i\hat L_i = \hat L_i\hat x_i =0 \;,\label{xdotl}\ee which implies 
\be   \hat A_i \hat A_i=\hat B_i\hat B_i \label{asqeqBsq}\ee
and hence  there is only one independent quadratic Casimir
 operator for  $so(4)$.

$ \hat A_i \hat A_i$, $\hat A_3$ and  $\hat B_3$ form a complete set of commuting operators, and so we can write down the three independent eigenvalue equations 
\beqa
\hat A_i\hat A_i\;|j,m_A,m_B> &=& j(j+1)\;|j,m_A,m_B>\cr & &\cr
\hat A_3\;|j,m_A,m_B> &=& m_A|j,m_A,m_B>\cr & &\cr
\hat B_3\;|j,m_A,m_B> &=& m_B\;|j,m_A,m_B> \;,\label{eigneqAB} \eeqa
where  $ m_A,m_B=-j,1-j,...,j\;,$   $j=0,\frac 12,1,\frac 32,...\;$. Since there is only one independent Casimir operator, we need only one index $j$ to label the irreducible representations of the  $so(4)$ algebra. The representations of 
 the Snyder algebra are reducible representations of  $so(4)$ because $\hat p_i$ does not commute with $ \hat A_i \hat A_i$.
 The  values of $j$ which occur in a given representation   of the Snyder algebra are given in the next section.  
The set of all eigenvalues  $\{|j,m_A,m_B>\}$  are also eigenstates of $\hat L_3$ and $\hat x_3$, 
\beqa \hat L_3\;|j,m_A,m_B>& =& (m_A+m_B)\;|j,m_A,m_B>\cr & &\cr\hat x_3\;|j,m_A,m_B>& =&\frac 1{\Lambda} \;(m_A-m_B)\;|j,m_A,m_B>
\label{xthreeigen}\eeqa 
The eigenvalues for $\hat x_3$  are evenly spaced.  The result also holds for   $\hat x_1$ and $\hat x_2$,  or any choice of Cartesian coordinates. Consequently, Snyder space corresponds to a cubical lattice with lattice size $\Lambda^{-1}$, where no two  directions are simultaneously measurable.   The  eigenvalues measured  in any particular direction are infinitely degenerate.  From (\ref{xthreeigen}), the set of degenerate eigenvectors associated with any particular eigenvalue  $n_3/{\Lambda},\;n_3=\; $integer, of $\hat x_3$  is \be\Bigl\{
|j,k,k-n_3>,\quad k\in {\rm integers},\quad j\ge |k| \;{\rm and }\;|k-n_3|\Bigr\}\label{degenpsigenv}\ee
  
It is not necessary to restrict to measurements along Cartesian directions.  For example,  by utilizing another  basis for $so(4)$, we can easily obtain the spectra for the radial coordinate.  This basis is associated with the sum of the two $SU(2)$ generators $\hat A_i+\hat B_i$, which is just the orbital angular momentum $L_i$.  The basis, which we denote by  $\{|j,\ell,m>_{\circ}\}$, diagonalizes  $\hat A_i\hat A_i$, 
   $\hat L_i\hat L_i$ and   $\hat L_3$, i.e.,
   \beqa
\hat L_i\hat L_i\;|j,\ell,m>_{\circ} &=& \ell(\ell+1)\;|j,\ell,m>_{\circ}\;\cr & &\cr
\hat L_3\;|j,\ell,m>_{\circ} &=& m\;|j,\ell,m>_{\circ} \;,\label{lthree} \eeqa
 $\ell$ and $m$ taking integer values, $\ell=2j, 2j-1,...,1,0$, and $m=m_A+m_B=-\ell,-\ell+1,...\ell\;$. $j(j+1)$ is once again the eigenvalue of $\hat A_i \hat A_i$.  
Of course, we also have
\be  (\hat L_1\pm i\hat L_2)\; |j,\ell,m>_{\circ}=\sqrt{\ell (\ell +1)-m(m\pm 1)}\;|j,\ell,m\pm 1>_{\circ}\label{lpmonsts}\ee
  The basis vectors $|j,\ell,m>_{\circ}$ are also eigenvectors of $\hat x_i\hat x_i$:
  \beqa
\hat x_i\hat x_i\;|j,\ell,m>_{\circ} &=&\frac { 4j(j+1)- \ell(\ell+1)}{\Lambda^2}\;|j,\ell,m>_{\circ} \label{xsqrigneq}\;,\eeqa
and so the radial coordinate takes the values $\sqrt{ 4j(j+1)- \ell(\ell+1)}/\Lambda$.
For a given $j$, it ranges from $\sqrt{2j}/\Lambda$ to $2\sqrt{j(j+1)}/\Lambda$ and  the eigenvalues  are $2\ell+1$ degenerate. Snyder space in this basis correspond to a set of concentric spheres, and as was  observed in \cite{Romero:2008jf},
the area of the spheres is quantized.  

\section{Momentum space}
\subsection{Differential representations}

\setcounter{equation}{0}

In the previous section, we only examined the algebra generated by $\hat x_i$ and $\hat L_i$.  Here we include the momentum operators $\hat p_i$.  The latter are simultaneously diagonalizable.  We denote their eigenvalues and eigenvectors by $p_i$ and $|\vec p>$, respectively, 
\be \hat p_i|\vec p>= p_i|\vec p>\ee The eigenvalues are continuous, and the set of all of them defines momentum  space.  In what follows, we   write down  differential representations of the $so(4)$ algebra on momentum space.  One such representation was given in Snyder's work\cite{Snyder:1946qz}. Here we extend Snyder's result to a one parameter family of differential representations.   An alternative representation is obtained by taking the Fourier transform,  which we briefly comment on at the end of this section.

We proceed by first writing down the deformation map from the Heisenberg algebra to the Snyder algebra.  The former is  generated by $\hat q_i$ and  $\hat p_i$, satisfying
\be [ \hat q_i,\hat p_j]=i\delta_{ij}\qquad  [ \hat q_i,\hat q_j]  =0\;,\label{cnclcmrls}\ee and of course $  [ \hat p_i,\hat p_j]  =0$.   
The  map relates  $\hat q_i$ to Snyder's position operators $\hat x_i$, and  is given by
\be \hat x_i = \hat q_i +\frac 1{2\Lambda^2}(\hat p_i\hat q_j \hat p_j +\hat p_j\hat q_j \hat p_i)  \;\label{qdarboux}\;,\ee 
while the momentum operators are unchanged by the map.

In the momentum representation, we replace $\hat q_i$ by $i\frac \partial{\partial p_i}$
and so the operators $\hat x_i $ and $\hat L_i$ can be realized by
\beqa \hat x_i&\rightarrow &i \frac{\partial}{\partial p_i} +\frac {i p_i}{\Lambda^2}\biggl(p_j  \frac{\partial}{\partial p_j }+2 \biggr)
\label{candifrep} \\&&\cr
\hat  L_i&\rightarrow & -{i}  \epsilon_{ijk} p_j \frac{\partial}{\partial p_k } \;,\label{dfrpxL} \eeqa  which  act on functions $\psi(\vec p)=<\vec p|\psi>$, where $|\psi>$ is a vector in the  Hilbert space for the Snyder algebra and we are using Dirac notation.  
From (\ref{dfrpxL}), the result found earlier  that $\ell$ is an integer means that wavefunctions in momentum space are single-valued.   The operators are symmetric for the scalar product $\; <\phi |\psi>=\int d^3p \;\phi(\vec p)^*\psi(\vec p)\;,\;$ provided  the functions $\phi$ and $\psi$ vanish sufficiently rapidly at spatial infinity in momentum space.  

 More generally, we can  preserve the Snyder algebra by adding a term proportional to $\hat p_i$ in (\ref{qdarboux}).
So we can generalize the differential operator  (\ref{candifrep}) to 
\beqa \hat x_i&\rightarrow &i \frac{\partial}{\partial p_i} +\frac{ ip_i}{\Lambda^2}\biggl( p_j  \frac{\partial}{\partial p_j }+\alpha \biggr)
\label{noncandifrep}\;,\eeqa
where we restrict $\alpha $ to the reals.  A value of $\alpha$ different from $2$  deforms the integration measure from $d^3p$ to  
\be d\mu(\vec p) =\frac{ d^3p}{(1+ \frac{ \vec p^2}{\Lambda^2})^{2-\alpha}} \label{measure}\;\ee 
The operators are now symmetric for the scalar product \be <\phi |\psi>=\int  d\mu(\vec p)\;\phi(\vec p)^*\psi(\vec p)\;,\label{sclrprdctpsc}\ee  for functions $\phi$ and $\psi$  satisfying asymptotic conditions which depend on $\alpha$.   $\;\int  d\mu(\vec p)\;|\vec p><\vec p|$ is the identity operator on the Hilbert space.    For $\psi(\vec p)$ to be normalizable it must go like $1/|\vec p|^w$, as $|\vec p|\rightarrow\infty$, where $w>\alpha-\frac 12$.
We  recover the trivial measure for  $\alpha=2$, while   $\alpha=0$  was  the   choice of  Snyder\cite{Snyder:1946qz}.\footnote{The $\alpha=0$ differential representation in \cite{Snyder:1946qz} was written down for the four-dimensional Minkowski version of (\ref{snydrsubalg}), and  the measure that appears there is ${ d^4p}/{(1+ \frac{ p^\mu p_\mu}{\Lambda^2})^{5/2}}$.}  For $\alpha=0$, normalizable  functions $\psi(\vec p)$ need not vanish as $|\vec p|\rightarrow\infty$.

Although in this article we shall rely exclusively on representations  in  momentum space, we here mention representations on the Fourier transform space.  The latter is spanned by functions $\tilde\psi(\vec q)=<\vec q|\psi>$,  $q_i$  and $|\vec q>$, respectively, denoting the eigenvalues and eigenvectors of   $\hat q_i$, satisfying (\ref{cnclcmrls}).  Here we represent $\hat p_i$ by $-i\frac \partial{\partial q_i}$,
and so from  (\ref{qdarboux}), $\hat x_i $ involves second order derivatives. $\hat x_i $ and $\hat L_i$ are realized by\footnote{  A similar representation was found in \cite{Hellund:1954zz}, however the differential operators there
were not symmetric.}
 \beqa \hat x_i& \rightarrow &q_i -\frac 1{\Lambda^2}
\biggl( 2\frac{\partial}{\partial q_i}  + q_j\frac{\partial^2}{\partial q_i\partial q_j}\biggr)
 \cr &&\cr
\hat  L_i&\rightarrow & -{i}  \epsilon_{ijk} q_j \frac{\partial}{\partial q_k } \label{qspcdfrpxL} \;\eeqa   The differential operators are symmetric  for the  scalar product which
 utilizes the  trivial measure $d^3q$; i.e., the scalar product between two functions $\tilde\phi$ and $\tilde\psi$ is
$<\tilde\phi |\tilde\psi>=\int d^3q \;\tilde\phi(\vec q)^*\tilde\psi(\vec q)$.

\subsection{Asymptotic conditions }

Here we argue that in order to define a consistent action of $SO(4)$, it is necessary to impose certain   asymptotic conditions in momentum space. The result is that momentum space is isomorphic to the $SO(3)$ group manifold. 
 For this,  we first write down the following map from the three-momentum to four operators $\hat P_A,\;A=1,2,3,4$,  according to
 \be \hat P_i =\frac{ \hat p_i}{\sqrt{\vec{\hat p}^2+\Lambda^2}}  \quad\qquad  \hat P_4 =\frac{ \Lambda}{\sqrt{\vec{\hat p}^2+\Lambda^2}}\label{coordnitize}\;\ee
Recalling that  $\hat L_{AB}$ in (\ref{dflab}) and 
 (\ref{so(4)algbr}) are the $SO(4)$ generators, one then finds that $\hat P_A$ rotates as  a four vector,
 \beqa [\hat L_{AB},\hat P_C]&=& i(\delta_{AC} \hat P_B -\delta_{BC} \hat P_A)
\eeqa
The four momentum operators are constrained by 
 \be \hat P_1^2+\hat P_2^2+\hat P_3^2+\hat P_4^2=\BI\;,\ee where $\BI$ is the identity, and so
  their eigenvalues  lie on $S^3$.   It corresponds to a slice of the four-dimensional de Sitter space discussed in \cite{Snyder:1946qz}.  
 The eigenvalues of $\hat P_A$ do not span all of $S^3$. Choosing $\Lambda >0$,  $\hat P_4$ has 
 positive-definite eigenvalues, and so  the coordinatization (\ref{coordnitize}) gives a restriction to one   hemisphere of $S^3$, while $\Lambda <0$ gives a restriction to the other hemisphere of $S^3$.    

Equivalently,  one can  define a $\Lambda-$dependent map $g^{(\Lambda)}$ from ${\mathbb{R}}^{3}=\{\vec p\}$ to $SU(2)$ according to 
\be g^{(\Lambda)}(\vec p)=\frac{1}{\sqrt{\vec{ p}^2+\Lambda^2}} \pmatrix{\Lambda +ip_3 &p_2+ip_1 \cr -p_2+ip_1 &\Lambda -ip_3\cr}\quad\in\quad SU(2)\label{lmdpndmp}\;,
\ee where, as before, $p_i$ denotes the eigenvalues of $\hat p_i$.
Applying the differential representations (\ref{dfrpxL}) and (\ref{noncandifrep}), one gets
\beqa \hat L_i\;g^{(\Lambda)}(\vec p)&=&-\frac 12\;[\sigma_i,g^{(\Lambda)}(\vec p)]  \cr &&\cr \hat x_i\;g^{(\Lambda)}(\vec p)&=&-\frac 1{2\Lambda}\;[\sigma_i,g^{(\Lambda)}(\vec p)]_+\;, \label{sofronsutwo}\eeqa 
where $\sigma_i$ are the Pauli matrices, $[\;,\;]_+$ denotes the anticommutator and we have chosen $\alpha=0$ for convenience.  It follows that $\hat A_i$ and $\hat B_i$ act, respectively, as left and right generators of $SU(2)$,
\beqa \hat A_i\;g^{(\Lambda)}(\vec p)&=&-\frac 12\;\sigma_i\;g^{(\Lambda)}(\vec p)  \cr &&\cr \hat B_i\;g^{(\Lambda)}(\vec p)&=&\frac 12\;g^{(\Lambda)}(\vec p)\;\sigma_i \label{ABonsutwo}\eeqa  
 The map (\ref{lmdpndmp}) when applied to all of momentum space, $ g^{(\Lambda)}\circ {\mathbb{R}}^{3}$, does not cover all of $SU(2)$.   Rather, assuming $\Lambda >0$, it is a restriction to $SU(2)$ matrices satisfying
\be {\rm Re}\;  g^{(\Lambda)}(\vec p)_{11}> 0,\;\;{\rm Re}\;  g^{(\Lambda)}(\vec p)_{22}\;\;>\;0\label{realitycndtn}\ee
 Furthermore,
  $ g^{(\Lambda)}\circ {\mathbb{R}}^{3}$,  for fixed $\Lambda $, is not invariant under  $SO(4)$ because points on one hemisphere  of $S^3$ can be rotated to the opposite hemisphere.  That is, the conditions  (\ref{realitycndtn}) are not preserved under  $SO(4)$. Thus the action of $\hat A_i$ and $\hat B_i$ in (\ref{ABonsutwo}) cannot be consistently exponentiated.

Let us next introduce the complementary map  $g^{(-\Lambda)}$  from ${\mathbb{R}}^{3}=\{\vec p\}$ to $SU(2)$.  It gives a restriction to $SU(2)$ matrices satisfying
\be {\rm Re}\;  g^{(-\Lambda)}(\vec p)_{11}<0,\;\;{\rm Re}\;  g^{(-\Lambda)}(\vec p)_{22}\;\;<\;0\ee
Then
 $  [g^{(\Lambda)}  \circ {\mathbb{R}}^{3}]\cup [g^{(-\Lambda)}\circ {\mathbb{R}}^{3}]$ spans all of $SU(2)$ and  is invariant under the  action of $SO(4)$.  For finite momentum, $|\vec p |<\infty$, the two maps  identify  each point $\vec p$ in  ${\mathbb{R}}^{3}$
with two  points in $SU(2)$.  Using
 \be g^{(-\Lambda)}(-\vec p)=-g^{(\Lambda)}(\vec p)\;,\label{idntfctnpnfnt}\ee the two maps are related by ${Z}_2=\{1,-1\}$.  There is thus a $2-1$ map from $SU(2)$ to $\{\vec p,\;|\vec p |<\infty\}$.
The restriction to finite momentum can be lifted upon  imposing appropriate asymptotic conditions in momentum space.  For (\ref{idntfctnpnfnt}) to hold
  as  $|\vec p|\rightarrow\infty$,  we need to identify opposite points at infinity,
\be\vec p\;\leftrightarrow\; -\vec p\;,\qquad {\rm as}\qquad|\vec p|\rightarrow\infty\label{asscond}\ee
These asymptotic conditions mean that momentum space is
 $SU(2)/Z_2=SO(3)$.   The $SO(3)$ matrices $\{R_{ij}(\vec p)\}$ are given explicitly by
\be\sigma_i R_{ij}(\vec p)= g^{(\Lambda)}(\vec p)\sigma_j g^{(\Lambda)}(\vec p)^\dagger\label{dfsothree}
\ee
Upon applying (\ref{lmdpndmp}),
\be  R(\vec p) =\frac 1{\vec p^2 +\Lambda^2}\left(
\begin{array}{lll}
 \Lambda^2+ p_1^2-p_2^2-p_3^2 & 2 ({p_1} {p_2}+\Lambda {p_3}) & 2 ({p_1} p_3- \Lambda {p_2} )\\
 2 ({p_1} p_2- \Lambda {p_3}) & \Lambda^2-p_1^2+p_2^2-p_3^2 & 2 (\Lambda {p_1}+{p_2} {p_3}) \\
 2 (\Lambda {p_2}+{p_1} {p_3}) & 2( {p_2} p_3- \Lambda {p_1}) & \Lambda^2-p_1^2-p_2^2+p_3^2
\end{array}
\right)\ee
From (\ref{sofronsutwo}), the action of $SO(4)$ generators on these matrices is given by
\beqa \hat L_i\;R_{jk}(\vec p)&=&-\;[T_i,R(\vec p)]_{jk}  \cr &&\cr \hat x_i\;R_{jk}(\vec p)&=&-\frac 1{\Lambda}\;\Bigl([T_i, R(\vec p)]_+\Bigr)_{jk}\;, \qquad (T_i)_{jk}=-i\;\epsilon_{ijk}\;,\eeqa 
where we have again chosen $\alpha=0$ for convenience.
Equivalently, $\hat A_i$ and $\hat B_i$ are, respectively, the left and right generators of $SO(3)$,
\beqa \hat A_i\;R_{jk}(\vec p)&=&-\;[T_iR(\vec p)]_{jk}  \cr &&\cr \hat B_i\;R_{jk}(\vec p)&=&[ R(\vec p) T_i]_{jk} \eeqa This action can be consistently exponentiated to $SO(3)\times SO(3)$ acting on momentum space by  left and right multiplication.

One can promote $R_{ij}$ to operator-valued matrix elements $\hat R_{ij}$.  For this we only need to replace $p_i$ by the operators $\hat p_i$ in the definition of the $SO(3)$ matrices in (\ref{dfsothree}).  The result is the set of operators $\hat R_{ij}=R_{ij}(\hat{\vec p})$, whose eigenvalues are $R_{ij}(\vec p)$
\be
\hat R_{ij}|\vec p>=R_{ij}({\vec p})|\vec p>
\ee  With the imposition of the asymptotic conditions (\ref{asscond}) on momentum space, the Snyder algebra  is  generated by $\hat A_i,\;\hat B_i$ and $\hat R_{ij}$. They satisfy commutation relations (\ref{ofralgbr}), along with
\beqa
[ \hat A_i,\hat R_{jk}]&=&-\;[T_i\hat R]_{jk}  \cr &&\cr [\hat B_i,\hat R_{jk}]&=&[ \hat R T_i]_{jk} \cr &&\cr [\hat R_{ij},\hat R_{k\ell}]&=&0 \label{sndrplstplg}
\eeqa 
This, along with (\ref{ofralgbr}), is an alternative way to write the three-dimensional Euclidean Snyder algebra, which takes into account the topology of momentum space.
In the next section we look at the different representations of this algebra.

\section{Wavefunctions on momentum space}

Here we find two distinct representations of the three-dimensional Euclidean Snyder algebra.  We examine them using  three different bases in the  subsections that follow.  The first deals directly with the algebra (\ref{ofralgbr}) and (\ref{sndrplstplg}), the second are momentum eigenfunctions of the radial coordinate $\sqrt{\hat x_i\hat x_i}$ and the third are momentum eigenfunctions of $\hat x_3$.

\setcounter{equation}{0}
\subsection{Two Hilbert spaces}

Multiple connectivity in a  classical theory implies that there are multiple quantizations of the system.\cite{Balachandran:1991zj} We can view  (\ref{ofralgbr}) and (\ref{sndrplstplg}) as resulting from  quantization on a  doubly connected momentum space.  This leads to two distinct representations.   We denote  the  corresponding  Hilbert spaces  by ${\cal H}_B$ and ${\cal H}_F$.  They are as follows:

The Hilbert space ${\cal H}_B$, consists of complex functions $\{\Phi_B,\Psi_B,...\}$ on $SO(3)=\{R\}$.  The scalar product between any two such functions $\Phi_B$ and $\Psi_B$ is an integral over $SO(3)$, 
\be   <\Phi_B |\Psi_B>=\int_{SO(3) } d\mu(R)_{SO(3)}\;\Phi_B(R )^*\Psi_B(R)\;,
\ee
where $d\mu(R)_{SO(3)}$ is the invariant measure on $SO(3)$.  From the Peter-Weyl theorem, any function
$\Psi_B$ in  ${\cal H}_B$ can be expanded in terms of the $(2j_B+1)
\times (2j_B+1)$ irreducible matrix representations $\{D^{j_B}(R),\; j_B=0,1,2,...\}$ of $SO(3)$, which serve as  basis functions,
\be \Psi_B(R)=\sum^\infty_{j_B=0}\;\;\sum_{m,n=-j_B}^{j_B}a^{j_B}_{mn}D^{j_B}_{mn}(R) \;, \ee $a^{j_B}_{mn}$ are constants.
$\hat A_i$ and $\hat B_i$ act on the irreducible  matrix representations according to
\beqa \hat A_i\;D^{j_B}_{mn}(R)&=&-\;[D^{j_B}(T_i)D^{j_B}(R)]_{mn}  \cr &&\cr \hat B_i\;D^{j_B}_{mn}(R)&=&[D^{j_B}( R)D^{j_B}( T_i)]_{mn}\;, \eeqa 
where $D^{j_B}(T_i)$ denote the $(2j_B+1)
\times (2j_B+1)$ matrix representations of the $SO(3)$ generators.  $\hat R_{ij}\;D^{j_B}_{mn}(R)$, for $j_B\ge 1$,  is a linear combination of  $D^{j_B+1}(R)$, $D^{j_B}(R)$  and  $D^{j_B-1}(R)$.  Applying the Casimir operator, one gets 
\be \hat A_i \hat A_i\;D^{j_B}_{mn}(R)=\hat B_i \hat B_i\;D^{j_B}_{mn}(R)=j_B(j_B+1)\;D^{j_B}_{mn}(R)\;,\ee
and so $j$ appearing  in (\ref{eigneqAB}) belongs to the set of all integers for the Hilbert space  ${\cal H}_B$.

The Hilbert space ${\cal H}_F$, consists of complex functions $\{\Phi_F,\Psi_F,...\}$ on $SU(2)=\{g\}$, with  scalar product  
\be   <\Phi_F |\Psi_F>=\int_{SU(2) } d\mu(g)_{SU(2)}\;\Phi_F(g )^*\Psi_F(g)\;,
\ee
where $d\mu(g)_{SU(2)}$ is the invariant measure on $SU(2)$.  The basis functions for  ${\cal H}_F$ are restricted to all half-integer irreducible matrix representations of $SU(2)$,  $\{{\tt D}^{j_F}(g),\; j_F=\frac 12,\frac 32,...\}$.
Thus any $\Psi_F$ in  ${\cal H}_F$ has the expansion 
\be \Psi_F(g)=\sum_{j_F=\frac12,\frac 32,...}\;\;\sum_{m,n=-j_F}^{j_F}{\tt a}^{j_F}_{mn}{\tt D}^{j_F}_{mn}(g) \;, \ee ${\tt a}^{j_F}_{mn}$ are constants.   $\hat A_i$ and $\hat B_i$  act on the irreducible  matrix representations according to
\beqa \hat A_i\;{\tt D}^{j_F}_{mn}(g)&=&-\;[{\tt D}^{j_F}(T_i){\tt D}^{j_F}(g)]_{mn}  \cr &&\cr \hat B_i\;{\tt D}^{j_F}_{mn}(g)&=&[{\tt D}^{j_F}( g){\tt D}^{j_F}( T_i)]_{mn}
\;, \eeqa 
where ${\tt D}^{j_F}(T_i)$ denote the $(2j_F+1)
\times (2j_F+1)$ matrix representations of the $SU(2)$ generators.  
Now \be \hat A_i \hat A_i\;{\tt D}^{j_F}_{mn}(g)=\hat B_i \hat B_i\;{\tt D}^{j_F}_{mn}(g)=j_F(j_F+1)\;{\tt D}^{j_F}_{mn}(g)\;,\ee
and  only half-integer values of $j=j_F$  occur in (\ref{eigneqAB}) for the Hilbert space  ${\cal H}_F$.

\subsection{Eigenfunctions of  $\hat x_i\hat x_i$, 
   $\hat L_i\hat L_i$ and   $\hat L_3$}

Here we write down  the momentum-dependent  eigenfunctions  $\phi_{j,\ell,m}(\vec p) = <\vec p|j,\ell,m>_{\circ} $ of  $\hat x_i\hat x_i$, 
   $\hat L_i\hat L_i$ and   $\hat L_3$.   (An alternative discussion of momentum eigenfunctions of the radial coordinate can be found in \cite{Mignemi:2011gr}.)  We show that these eigenfunctions   are related  to the  spherical harmonics of $S^3$  - restricted to one hemisphere $S^3_+$. As in the previous subsection, we find  two distinct Hilbert spaces, one consisting of $\phi_{j,\ell,m}(\vec p)$ with $j$ integer and the other $j$ half-integer.  Here we do not make any initial  assumptions on the domain of the wavefunctions, such as (\ref{asscond}).
For generality,  we drop the restriction to $\alpha =0$, which was  made in the previous subsection.
Using (\ref{noncandifrep}), the differential representation of  $\hat x_i\hat x_i$ is given by 
\be -\Lambda^2 \hat x_i\hat x_i  \rightarrow  -\frac {\hat L_i\hat L_i}{\rho^2} \;+\;(1+\rho^2)^2 \Bigl(\partial_\rho^2 +\frac 2\rho \partial_\rho\Bigl)\; +\;\alpha \Bigl[2(1+\rho^2)\rho\partial_\rho + (1+\alpha)\rho^2+3 \Bigr]\;,
\ee 
where $ \rho$ is the rescaled radial component of the momentum, $\rho= {|\vec p|}/\Lambda$.  Its eigenfunctions are proportional to the spherical harmonics $Y^\ell_m(\theta,\phi)$ on $S^2$,
\be \phi_{j,\ell,m}(\vec p)\; = \;\frac 1\rho u_{j,l}(\rho)\; Y^\ell_m(\theta,\phi) \;,\ee where $\theta$ and $\phi$ are spherical angles in momentum space.  From the eigenvalue equation (\ref{xsqrigneq}) we get  the following differential equation   for   the  function $u_{j,\ell}(\rho)$ 
\be \biggl\{(1+\rho^2) \frac{\partial^2}{\partial\rho^2} \; +\;2\alpha \rho\frac{\partial}{\partial\rho}  \; +\;\frac {4j(j+1)+\alpha\Bigl(1+\rho^2(\alpha- 1)\Bigr)}{1+\rho^2}\; -\;\frac{\ell(\ell +1)}{\rho^2}\biggl\}u_{j,l}(\rho)\;=\;0  \label{dif3eqfu}\ee
Solutions for $u_{j,\ell}(\rho)$ 
 involve  Gegenbauer polynomials $C^{(b)}_n$, where $n$ is a nonnegative integer and $b>\frac 12$. 
They are \be u_{j,\ell}(\rho)=\;
\;{\cal N}_{j,\ell}\; \sin^{\ell+1}\chi\cos^{\alpha -1}\chi\; C^{(\ell+1)}_{2j-\ell}(\cos\chi) \;,\label{ujleignfnphi}\ee
where  ${\cal N}_{j,\ell}$ are  normalization constants.  The angle $\chi$ is defined by
\be \tan\chi= \rho \;,\label{rhotochi}\ee and  runs from $0$ to $\pi/2$.  Then  the  eigenfunctions
 of  $ \phi_{j,\ell,m}(\vec p)$
are   
\be \phi_{j,\ell,m}(\vec p)\; = \;{\cal N}_{j,\ell}\; \sin^\ell\chi\;\cos^{\alpha }\chi \;C^{(\ell+1)}_{2j-\ell}(\cos\chi) \;Y^\ell_m(\theta,\phi)\;\label{eignfnphi}\ee
Using the measure (\ref{measure}), the norm of $ \phi_{j,\ell,m}(\vec p)$ is finite due to the ultraviolet scale $\Lambda$.  Furthermore, it is   independent of the parameter $\alpha$, as  the $\alpha$ dependence in $ \phi_{j,\ell,m}(\vec p)$ is canceled out by the $\alpha$ dependence in the measure.

The eigenfunctions $\phi_{j,\ell,m}(\vec p)$ are related to  spherical harmonics on $S^3$. If  we set $\alpha=0$,  they are in fact identical to  the  spherical harmonics, up to a normalization factor, and are obtainable from $O(4)$ representation matrices.\cite{tolman}  However, their domain is not all of $S^3$.  To see this we can embed  the three-sphere in ${\mathbb{R}}^{4}$, by defining 
 \beqa P_1&=&\sin \chi \sin\theta \cos\phi\cr
 P_2&=&\sin \chi \sin\theta \sin\phi\cr
 P_3&=&\sin \chi \cos\theta\cr
 P_4&=&\cos \chi  \label{crtntrmsph}\eeqa
 It follows that  $  P_1^2+ P_2^2+ P_3^2+ P_4^2=1$, and like  the eigenvalues of the operators $\hat P_A$ defined in (\ref{coordnitize}), $P_A$ span a hemisphere of $S^3$.  As before, $P_4\ge0$, since 
 $0\le\chi\le\frac \pi 2$.  The restriction of the domain of the spherical harmonics to the half-sphere $S^3_+$ affects their normalization.  More significantly, it affects their completeness relations, as we discuss below. 

Demanding that $ \phi_{j,\ell,m}(\vec p)$ are orthonormal, 
 \be <\phi_{j,\ell,m}|\phi_{j',\ell',m'}>\;=\;\int d\mu(\vec p) \;\phi_{j,\ell,m}(\vec p)^*\phi_{j'\ell',m'}(\vec p) \; =\;\delta_{j,j'}\delta_{\ell,\ell'}\delta_{m,m'}\;,\ee leads to the following conditions on the Gegenbauer polynomials
\be \Lambda^3 {\cal N}_{j,\ell}^*{\cal N}_{j',\ell'}\int_0^{\frac \pi 2} d\chi \;\sin^{2\ell +2}\chi \;C^{(\ell+1)}_{2j-\ell}(\cos\chi)\; C^{(\ell+1)}_{2j'-\ell}(\cos\chi) \;=\; \delta_{j,j'} \;\label{ggnbrhlflin}\;,\ee
using the measure (\ref{measure}).
  Gegenbauer polynomials $\{C^{(\ell+1)}_{n}(\xi)\}$  are standardly normalized  over the domain  $-1\le \xi \le 1$, or equivalently $0\le \chi\le \pi$, rather than  $0\le \chi\le \pi/2$.  The standard normalization condition is 
\be \int_0^{ \pi } d\chi \;\sin^{2\ell +2}\chi \;C^{(\ell+1)}_{n}(\cos\chi)\; C^{(\ell+1)}_{n'}(\cos\chi) \;=\; \frac{\pi \;(n+2\ell +1)!}{2^{2\ell +1}\; n!\; (n+\ell+1)(\ell !)^2}\;\delta_{n,n'}\label{ggnbrortho}\ee
In order to relate  this to (\ref{ggnbrhlflin}), we can use the property
\be 
 C^{(\ell+1)}_{n}(\xi)=(-1)^n C^{(\ell+1)}_{n}(-\xi)\label{parity}\ee    Then   $\{C^{(\ell+1)}_{n}(\xi), n\;$even$\}$ and 
 $\{C^{(\ell+1)}_{n}(\xi), n\;$odd$\}$ form two sets of orthogonal  polynomials over the half-domain, $0\le \xi \le 1$, or equivalently  $0\le \chi\le \pi/2$.  From (\ref{ggnbrhlflin}) and (\ref{ggnbrortho}), the normalization constants are given by 
 \be  | {\cal N}_{j,\ell}|^2 =\frac{2^{2\ell+2}(2j+1)(2j-\ell)!(\ell!)^2}{\Lambda^3 \pi (2j+\ell+1)!}\ee
We note that   the set  $\{C^{(\ell+1)}_{n}(\xi), n\;$even$\}$ is {\it not} orthogonal to   $\{C^{(\ell+1)}_{n}(\xi), n\;$odd$\}$ over the half-domain, $0\le \xi \le 1$.  Thus there are two distinct sets of orthonormal polynomials.  Using $n=2j-\ell$,  they correspond to either $j$ equal to an integer or $j$ equal to a half-integer, and for any given value of $\ell$.  There are then  two distinct sets of orthonormal  eigenfunctions $ \phi_{j,\ell,m}(\vec p)$,
and two distinct sets of spherical harmonics on  $S^3_+$.  As in the previous subsection,  we find that   there are two  representations of the Snyder algebra, ${\cal H}_B$ and ${\cal H}_F$, the former associated with all integer values of $j$ and the latter associated with all half-integer values of $j$.   

Unlike  the  derivation in subsection 4.1, here we did not a priori make any assumptions about the topology of momentum space, such as   (\ref{asscond}), which identifies opposite points at infinity.  The two bases of eigenfunctions which result here are distinguished by their asymptotic properties.  Restricting to $\alpha=0$, we get
\be \phi_{j,\ell,m}(\vec p)\sim C^{(\ell+1)}_{2j-\ell}(0) \;Y^\ell_m(\theta,\phi)\qquad {\rm as }\quad |\vec p| \rightarrow\infty\label{astmptphi}\ee  From (\ref{parity}), $C^{(\ell+1)}_{2j-\ell}(0) $ vanishes when $2j-\ell$ equals an odd integer.  Using this and the well known property
 \be Y^\ell_m(\pi-\theta,\pi+\phi)=(-1)^\ell\; Y^\ell_m(\theta,\phi)\;,\ee
it follows that the asymptotic expression (\ref{astmptphi}) has even parity when $j$ is an integer and odd parity when $j$ is a half-integer.  Trivial examples of this are  the zero angular momentum, parity even, eigenfunctions 
\be \phi_{j,0,0}(\vec p)= \frac{\sin(1+2j) \chi \; } {\Lambda^{3/2}\pi\; \sin\chi} \label{zamsts1}\;,\ee  again assuming $\alpha=0$, which vanish as $|\vec p|\rightarrow\infty$  when $j$ is half-integer. In general,  wavefunctions $\Psi_B$ in ${\cal H}_B$ satisfy
\be\Psi_B(\vec p)=\Psi_B( -\vec p)\;,\qquad {\rm as}\qquad|\vec p|\rightarrow\infty\;,\ee
while  $\Psi_F$ in ${\cal H}_F$ satisfy
\be\Psi_F(\vec p)=-\Psi_F( -\vec p)\;,\qquad {\rm as}\qquad|\vec p|\rightarrow\infty\ee

The eigenfunctions (\ref{eignfnphi}) provide  a  transform from momentum space to   discrete position space, which in this basis is  composed of concentric spheres of radii equal to $\sqrt{ 4j(j+1)- \ell(\ell+1)}/\Lambda$.  If $\Psi(\vec p)$ denotes a wavefunction in the former space  and  $\Psi^\circ_{j,\ell,m}$ is the corresponding wavefunction on the discrete space, then the transform and its inverse  are given by
\beqa \Psi^\circ_{j,\ell,m}&=&\int d\mu(\vec p) \;\phi_{j,\ell,m}(\vec p)^*\Psi(\vec p)\label{psicirc}
\\ & &\cr
\Psi(\vec p)
&=&\sum_{\left.\matrix{j=0,1,2,...\cr {\rm  or}\cr j= \frac 12, \frac 32,\frac 52,...\cr}
\right.} \sum_{\ell=0}^{2j} \sum_{m=-\ell}^\ell\phi_{j,\ell,m}(\vec p)\;\Psi^\circ_{j,\ell,m}\;\label{fltrnsfcsp}
\eeqa
The sum over integer $j$ is for Hilbert space  ${\cal H}_B$ and half-integer $j$ is for  ${\cal H}_F$.
In  Appendix B, we give an example of such a transform, which can be performed exactly. 

Finally, in addition to (\ref{ujleignfnphi}), there are another set of solutions for $  u_{j,l}(\rho)$ in (\ref{dif3eqfu}).  They can be expressed in terms of hypergeometric functions ${}  _2F_1$ according to
\be \frac{\rho^{-\ell} }{\left(1+\rho^2\right)^{j+\frac{\alpha }{2}}} \;
   _2F_1\left(-j-\frac{\ell}{2}-\frac{1}{2},-j-\frac{\ell}{2};\frac{1}{2}-\ell;-\rho^2\right)
 \ee
They lead to a complementary set of eigenfunctions
$ \{\phi'_{j,\ell,m}(\vec p)\}$ of   $\hat x_i\hat x_i $,     $\hat L_i\hat L_i$, and   $\hat L_3$, which unlike (\ref{ujleignfnphi}),  are singular at the origin in momentum space.  The singularity is integrable only for the case of zero angular momentum, where   the eigenfunction is given by
\be \phi'_{j,0,0}(\vec p)= \frac{\cos(1+2j) \chi \;} {\Lambda^{3/2}\pi\; \sin\chi} \label{zamsts2}\;,\ee and we
again take $\alpha=0.$  The solutions
(\ref{zamsts1}) and (\ref{zamsts2}) comprise the spherically symmetric waves for the system.

\subsection{Eigenfunctions of  $\hat A_i\hat A_i$, 
   $\hat x_3$ and   $\hat L_3$}

In the previous section we found eigenfunctions of the radial coordinate operator.  Here we examine eigenfunctions of $\hat x_3$.  More precisely, we study the momentum-dependent   basis functions  associated with the eigenvectors $\{|j,m_A,m_B>\} $ of Sec 2.2. We denote these basis functions by $\eta _{j,m_A,m_B}(\vec p) = <\vec p|j,m_A,m_B> $ .  They are obtained from  $\phi_{j,\ell,m}(\vec p)$ in (\ref{eignfnphi}) by a change of basis 
\be \phi_{j,\ell,m}(\vec p) =\sum_{m_A=-j}^{j}<j,j;m_A,m-m_A|\ell,m> \eta_{j,m_A,m-m_A}(\vec p)\;,\ee
where $<j,j;m_A,m_B|\ell,m>$ are Clebsch-Gordan coefficients.  
Expressions for  $\eta _{j,m_A,m_B}(\vec p)$ can be given in terms of the cylindrical variables $R_p,\;\phi_p,\;p_3$, [where $p_1=R_p\cos\phi_p,\; p_2=R_p\sin\phi_p$ and $0\le \phi_p<2\pi$].  The eigenfunctions have the general form
\be \eta_{j,m_A,m_B}(\vec p) =f_{j,m_A,m_B}(R_p,p_3)\; e^{i(m_A+m_B)\phi_p} \;, \ee where
$f_{j,m_A,m_B}(R_p,p_3)$ are eigenfunctions of $\hat A_i\hat A_i$ and 
   $\hat x_3$.
We can obtain the precise form of $f_{j,m_A,m_B}(R_p,p_3)$  for the special case of
zero angular momentum in the third direction. Then, $\eta_{j,m_A,-m_A}(\vec p)= f_{j,m_A,-m_A}(R_p,p_3)$,
 corresponding to plane waves in the third direction.   From the eigenvalue equation \be\hat x_3\;\eta_{j,m_A,-m_A}(\vec p)=2m_A\;\eta_{j,m_A,-m_A}(\vec p)\;,\ee we get
\be
 \eta_{j,m_A,-m_A}(\vec p) = {e^{-2i\;m_A \,\tan^{-1}( \frac{{p_3}}{\Lambda })}}\;\;{{\cal F}_{j,m_A}\Bigl(\frac{{{\vec p}}^2 +{\Lambda }^2}{{{p_3}}^2 + {\Lambda }^2}\Bigr)}
\;,\ee where we used the differential representation (\ref{noncandifrep}) with  $\alpha=0$.  The functions  ${\cal F}_{j,m_A}(\zeta)$ are determined from the remaining  eigenvalue equation  \be\hat A_i \hat A _i\;\eta_{j,m_A,-m_A}(\vec p)=j(j+1)\eta_{j,m_A,-m_A}(\vec p)\;,\ee  which leads to
\be \zeta^2 \frac{d}{d\zeta}\Bigl((\zeta-1) \frac{d}{d\zeta}\Bigr){\cal F}_{j,m_A}(\zeta) +(j^2+j-{m_A}^2 \zeta){\cal F}_{j,m_A}(\zeta) =0 
\ee Solutions for ${\cal F}_{j,m_A}(\zeta)$ can be expressed in terms of hypergeometric functions. Up to a  normalization, ${\cal F}_{j,m_A}(\zeta)$ is 
\be \zeta^{-j} \; _2F_1\left(-j-{m_A},{m_A}-j;-2 j;\zeta\right)
\ee Additional solutions are   
\be \zeta^{j+1} \; _2F_1\left(j-{m_A}+1,j+{m_A}+1;2 j+2;\zeta \right)\;,
\ee which are singular at $\zeta =1$.  Using $\zeta =({{\vec p}}^2 +{\Lambda }^2)/({p_3}^2 + {\Lambda }^2)$, the latter are divergent along $p_3$-axis.

The eigenfunctions $\eta_{j,m_A,m_B}(\vec p)$ provide  a  transform from momentum space to the spatial lattice (with only  $x_3$  determined).  If $\Psi(\vec p)$ is a wavefunction in the former space  and  $ \Psi^{\#}_{j,m_A,m_B}$ is a wavefunction on the lattice, then they are related by
\beqa  \Psi^{\#}_{j,m_A,m_B}&=&\int d\mu(\vec p) \;\eta_{j,m_A,m_B}(\vec p)^*\;\Psi(\vec p)
\\ & &\cr
\Psi(\vec p)
&=&\sum_{\left.\matrix{j=0,1,2,...\cr {\rm  or}\cr j= \frac 12, \frac 32,\frac 52,...\cr}
\right.} \sum_{m_A,m_B=-j}^j \eta_{j,m_A,m_B}(\vec p)\; \Psi^{\#}_{j,m_A,m_B}\;
\eeqa
Once again, 
the sum over integer $j$ is for Hilbert space  ${\cal H}_B$ and half-integer $j$ is for  ${\cal H}_F$.

\section{Rotations and Translations}

\setcounter{equation}{0}

It is now straightforward to write down the unitary action of the rotation and translation group on Snyder space.  They are generated by  $\hat L_i$ and  $\hat p_i$, respectively.  The latter is not responsible for discrete translations from  one point on the spatial lattice to a neighboring point. Rather, $\hat p_i$ generate continuous translations in the basis $\{|\vec q>\}$ which diagonalizes the operators $\hat q_i$ conjugate to $\hat p_i$. [Cf. eq. (\ref{cnclcmrls}).]  Differential representations associated with this basis were given in eq. (\ref{qspcdfrpxL}).

In the basis $\{|j,\ell,m>_{\circ}\}$, Snyder space corresponds to the set of concentric spheres with radii  equal to $\sqrt{ 4j(j+1)- \ell(\ell+1)}/\Lambda$.  $m=-\ell,...\ell$ is a degeneracy index. A rotation by $\vec \theta$ is given by
\be  e^{i\theta_i \hat L_i}|j,\ell,m>_{\circ}\;=\;\sum_{m'=-\ell}^\ell\; D^\ell_{m',m}(\vec\theta)\,|j,\ell,m'>_{\circ}\;,\ee
where $D^\ell$ are $SO(3)$ matrix representations, 
$ D^\ell_{m',m}(\vec\theta) =<\ell,m'| e^{i\theta_i \hat L_i}|\ell,m>$ and 
 $|\ell,m>$ are the usual angular momentum eigenstates. A translation by $\vec a$ is given by
\beqa e^{ia_i \hat p_i}|j,\ell,m>_{\circ}&=&\cr & &\cr \sum_{\left.\matrix{j'=0,1,2,...\cr {\rm  or}\cr j'= \frac 12, \frac 32,\frac 52,...\cr}
\right.}\sum_{\ell'=0}^{2j'}
\;\sum_{m'=-\ell'}^{\ell'}
 \;\int d\mu(\vec p) & &|j',\ell',m'>_{\circ}\; e^{ia_i  p_i}\;\phi _{j',\ell',m'}(\vec p)^*\;\phi _{j,\ell,m}(\vec p)\;, \label{unitrytrnsltn}\cr & &\eeqa 
and one sums over integer (half-integer) values of   $j'$  when $j$ is  integer (half-integer), because the translation operator does not take states out of the Hilbert space. 

In the basis $\{|j,m_A,m_B>\}$, Snyder space corresponds to 
a lattice with the eigenvalues of the third coordinate given by $(m_A-m_B)/\Lambda$.
Using the Clebsch-Gordan coefficients,  rotations act according to
\beqa  e^{i\theta_i \hat L_i}|j,m_A,m_B>&=&\cr & &\cr\sum_{\ell=0}^{2j}\;
\sum_{m=-\ell}^{\ell}\;
\sum_{m'_A,m'_B=-j}^j  & &|j,m'_A,m'_B><j,j;m'_A,m'_B|\ell,m>\;\times\cr & &\cr& &<j,j;m_A,m_B|\ell,m_A+m_B>^*\, D^\ell_{m,m_A+m_B}(\vec\theta)\;, \cr
& &\eeqa
while translations are given by \beqa e^{ia_i \hat p_i}|j,m_A,m_B>&=&\cr & &\cr\sum_{\left.\matrix{j'=0,1,2,...\cr {\rm  or}\cr j'= \frac 12, \frac 32,\frac 52,...\cr}
\right.} \sum_{m'_A,m'_B=-j'}^{j'}
 \;\int d\mu(\vec p) & &|j',m'_A,m'_B> e^{ia_i  p_i}\;\eta _{j',m'_A,m'_B}(\vec p)^*\;\eta _{j,m_A,m_B}(\vec p)\;, \label{unitrytrnsltn}\cr & &\eeqa 
Again,  one sums over integer (half-integer) values of   $j'$  when $j$ is  integer (half-integer). As stated above, $e^{ia_i \hat p_i}$    does not, in general, map one point on  the spatial lattice to another point.    To hop  from one eigenvalue of $\hat x_3$ to another, one needs an operator that maps $(m_A,m_B)$ to $(m'_A,m'_B)$, with $m_A-m_B\ne m'_A-m'_B$.  Such an operator need not be unitary; Examples are the raising and lowering operators $\hat A_1\pm i \hat A_2$ and $\hat B_1\pm i \hat B_2$.

\section{Discussion}

\setcounter{equation}{0}
We have obtained  basis functions for the Snyder algebra and have argued that they span two distinct Hilbert spaces, ${\cal H}_B$, and  ${\cal H}_F$.  The Hilbert spaces are associated with  different lattice structures.  Two different lattice structures were also noted in the case of particles in $2+1$ gravity and  were referred to as bosonic and fermionic.\cite{'tHooft:1996uc} In our case, the lattices are distinguished by an $SU(2)$ quantum number, which takes integer values for ${\cal H}_B$  and half-integer values for ${\cal H}_F$.  In analogy with spin $SU(2)$, there may exist a lattice-statistics connection, whereby bosons are associated with ${\cal H}_B$ and and fermions  are associated with ${\cal H}_F$.  Also, a supersymmetric generalization of this system may lead to ${\cal H}_B$ and  ${\cal H}_F$ coexisting in a graded space.

We showed that rotations and translations can be unitarily implemented on the lattice structures.  In generalizing to the Poincar\'e group for relativistic systems (or Galilei group for nonrelativistic systems), we need to re-introduce the notions of time and energy, which were associated with independent  generators in  Snyder's  algebra.\cite{Snyder:1946qz}
Snyder's time operator has a continuous spectrum,  and its commutation relations  were postulated in order to maintain Lorentz invariance.
Time operators can be a source of additional conceptual problems in quantum mechanics.  Alternatively, one can introduce the notion of  time through particle dynamics.
  Previous articles have addressed the issue of particle dynamics in Snyder space-time,\cite{Jaroszkiewicz}, \cite{Mignemi:2011gr}  a number of which  have focused on the problem of deriving the Snyder algebra starting from an action principle.\cite{Romero:2004er},\cite{Romero:2006pe},\cite{Banerjee:2006wf},\cite{Chatterjee:2008bp},\cite{Stern:2010ri} In a forthcoming work,\cite{leilu} we shall  examine the motion of a relativistic  particle on the three-dimensional Euclidean Snyder space discussed here.  We shall show that, `time' can be either  identified with a real number $\lambda$ which parametrizes the particle evolution, or it can be some function of $\lambda $ and the fundamental operators, $\hat x_i$ and $\hat p_i$.  In the latter,   time is an $\lambda$-dependent operator whose commutation relations can be derived from   (\ref{snydrsubalg}).
The resulting algebra is not Lorentz covariant.  Nevertheless, the time operator can be used to construct Lorentz boost operators ,  from which we can then recover the Lorentz algebra.
Upon including the free particle Hamiltonian operator, one recovers the 
Poincar\'e algebra, and so the  full Poincar\'e group can be unitarily implemented on the lattice.

Finally, there have been various attempts to write down field theory on Snyder space with the goal of having a consistent Lorentz invariant noncommutative quantum field theory\cite{Battisti:2010sr},\cite{Girelli:2010wi}.  The approach considered so far  is to try to write down star product representations of the Snyder algebra on a commutative space-time manifold.  The proposed star products have introduced some confusion, in that they are either nonassociative  or lead to a deformation of the Poincar\'e symmetries.\cite{Battisti:2010sr},\cite{Girelli:2010wi}   Since Snyder space is, in fact, a lattice,  a more appropriate way to proceed with the  second quantization of the theory may be to consider  lattice field theory.   Unlike usual lattice field theory, here fields can be defined in only one direction of the spatial lattice; this would correspond, for example, to the radial direction if we use the basis $\{|j,\ell,m>_{\circ}\}$, or the third Cartesian direction if we use the basis $\{|j,m_A,m_B>\}$.
Assuming the latter, fields are  a function of quantum numbers $j,m_A$ and $m_B$, and some continuous evolution parameter, say $\lambda$. It can also depend on spin quantum numbers.  (As stated above, the system, here expressed in terms of fields,  should carry representations of the Poincar\'e group.)
In the simplest case of a real scalar field $\Phi(\lambda)_{j,m_A,m_B}$, we can write
\be   \Phi(\lambda)_{j,m_A,m_B}=\int d\mu(\vec p)\;\Bigl( {\bf a}(\vec p)   \; \eta_{j,m_A,m_B}(\vec p)\;e^{-i\lambda H} + {\bf a}(\vec p)^\dagger   \; \eta_{j,m_A,m_B}(\vec p)^*\;e^{i\lambda H}\Bigr)\;,
\ee  where $H$ generates evolution in $\lambda$, and we introduced particle creation (annihilation) operators $ {\bf a}(\vec p)^\dagger\;\Bigl( {\bf a}(\vec p) \Bigr)$.  The field action $ {\cal S}[\Phi]$ involves  an integral over $\lambda$, as well as a  sum over  the lattice, 
\be  {\cal S}[\Phi] =\int d\lambda\; \sum_{{\cal H}_B\;{\rm  or}\; {\cal H}_F}\;{\cal L}[\Phi(\lambda)]\;,\ee where 
${\cal L}[\Phi(\lambda)]$ is the field Lagrangian.   If the lattice-statistics connection mentioned above applies, then the sum is over the Hilbert space ${\cal H}_B$ for bosonic fields and  ${\cal H}_F$ for fermionic fields.   A nontrivial problem is then to find a Lagrangian on the  lattice which gives  relativistic invariant dynamics and reduces to a familiar model in the limit of zero lattice spacing $\Lambda\rightarrow\infty$. 

\bigskip

{\Large {\bf Acknowledgments} }

\noindent
 We are very grateful for useful discussions with L. Dabrowski, G. Landi, T. Kephart and G. Piacitelli.  A. S. thanks  L. Dabrowski and G. Landi for their hospitality during a visit to SISSA.
This work was supported in part by the DOE,
Grant No. DE-FG02-10ER41714.

\bigskip
\appendix
\appendice{\Large{\bf $\quad$ Sample wavepackets}}

Here we construct  normalizable wavepackets in momentum space which can be explicitly transformed to the discrete position space.  The transformed wavepackets are  functions on the set of concentric spheres with radii
$\sqrt{ 4j(j+1)- \ell(\ell+1)}/\Lambda$. For the examples which follow, we can carry out the integral in (\ref{psicirc}) 
analytically.
  The transformation can be performed explicitly thanks to  the following well known  identity for Gegenbauer polynomials: 
\be \frac 1{(1- 2\kappa \cos\chi + \kappa^2)^{\ell+1}} = \sum_{n=0}^\infty \;C^{(\ell+1)}_{n}(\cos\chi)\,\kappa^n \;,\label{ggnbridnte} \ee where $\kappa$ is in general a complex parameter with $|\kappa|<1$. 
 From subsection 4.2, the  Gegenbauer polynomials which appear in the eigenfunctions $\phi_{j,\ell,m}(\vec p)$ 
 are restricted to the half-interval, $0\le \chi\le \pi/2$, and they form  two  orthogonal sets, those with $n$  even or $n$  odd.
From 
(\ref{eignfnphi}), $n$ should be identified with $2j-\ell$.  Using the property (\ref{parity}), one  can obtain two additional identities from (\ref{ggnbridnte}) which involve sums either  over all even $n$ or all odd $n$:
\beqa f^{(\ell,\kappa)}_\pm(\chi)&=&\frac 1{(1+ 2\kappa \cos\chi + \kappa^2)^{\ell+1}}\pm \frac 1{(1- 2\kappa \cos\chi + \kappa^2)^{\ell+1}} \cr && \cr &= &\sum_{ n=  \Bigl\{ {} ^{{}^ {\mbox{\tiny 0,2,4,...}}}_{{}_{\mbox{\tiny 1,3,5,...}}} \Bigr\}} \;C^{(\ell+1)}_{n}(\cos\chi)\,\kappa^n \;\label{twfts} \eeqa  The function $ f^{(\ell,\kappa)}_+$ is not normalizable, i.e., $\int_0^{\frac \pi 2} d\chi \;\sin^{2\ell +2}\chi \;|f^{(\ell,\kappa)}_+(\chi)|^2$ is not finite, and so we restrict to $ f^{(\ell,\kappa)}_-$ and sum over odd  
$n$.  For any given $\ell, m$ and $\kappa$,   we can then construct the normalizable momentum-dependent wavepacket 
\beqa \Psi^{(\ell,m,\kappa)}(\vec p)&=& {\cal A}(\kappa)_{\ell,m}\; f^{(\ell,\kappa)}_-(\chi)\; \sin^\ell\chi\; \;Y^\ell_m(\theta,\phi)\;,\eeqa  
where the momentum is defined in terms of angles $\chi,\theta$ and $\phi$ according to (\ref{crtntrmsph}) and $ {{\cal A}(\kappa)_{\ell,m}}$ is a normalization factor.  It can be expanded in terms of basis functions $\phi_{j,\ell,m}$ defined in (\ref{eignfnphi}), 
\beqa \Psi^{(\ell,m,\kappa)}(\vec p) &=& {\cal A}(\kappa)_{\ell,m}\sum_{2j-\ell=1,3,5...}\frac{\kappa^{2j-\ell}}{{\cal N}_{j,\ell}}\;\phi_{j,\ell,m}(\vec p)\;,\eeqa  
and we have chosen $\alpha = 0$ for convenience.
In comparing with 
(\ref{fltrnsfcsp}), the corresponding wavefunction $ \Psi^{\circ(\ell,m,\kappa)}$ in position space is given by
\be  \Psi^{\circ(\ell,m,\kappa)}_{j,\ell',m'}=\frac {{\cal A}(\kappa)_{\ell,m}}{{\cal N}_{j,\ell}}{\kappa^{2j-\ell}}\delta_{\ell,\ell'}\;\delta_{m,m'}  \;\ee 
As an example, we can consider spherically symmetric wavepacket, $\ell=m=0$, 
\beqa \Psi^{(0,0,\kappa)}(\vec p)
&=& \frac{\sqrt{1-\kappa^4}}{(\Lambda \pi)^{3/4}} \;\frac{2\cos\chi}{1-2\kappa^2 \cos{2\chi} +\kappa^4}  \label{bateenine} \eeqa Its transform  to position space is
\beqa  \Psi^{\circ(0,0,\kappa)}_{j,\ell',m'}&=&\frac {\sqrt{1-\kappa^4}}{2}{\kappa^{2j-1}}\delta_{\ell',0}\;\delta_{m',0} 
\;\eeqa
From (\ref{bateenine}) and  the differential representation (\ref{noncandifrep}), one can compute the mean values of $\vec x^2$ and $\vec p^2$ for these wavepackets.  The result is 
\beqa <\vec x^2>_{(0,0,\kappa)} &=& \frac{3 + 6\kappa^4  -\kappa^8}{\Lambda^2(1-\kappa^4)^2}\cr & &\cr<\vec p^2>_{(0,0,\kappa)} &=& \Lambda^2\frac{(1-\kappa^2)(3+\kappa^2)}{(1+\kappa^2)^2 }\qquad\qquad \eeqa

\appendice{\Large{\bf $\quad$Comparison with the Hydrogen atom}}

There are several similarities, along with some interesting differences, between  the  three-dimensional (Euclidean) Snyder algebra (\ref{snydrsubalg})  and  the algebra of the hydrogen atom, which we point out in this appendix. 
Both systems contain an $so(4)$ subalgebra and the representations which appear in the quantum theory are infinite-dimensional and  reducible. However, their action on momentum space is different for the two systems.

We recall some features of the   $so(4)$ algebra for the hydrogen atom, first in the classical theory.  The group is generated by the angular momentum $L^{\tt H}_i$ and
 \be   x^{\tt H}_i =\frac 1{\sqrt{-H}}\; K^{\tt H}_i\;,\ee where
$H$ is the Hamiltonian 
\beqa 
H&=& \frac{p^{\tt H}_i p^{\tt H}_i}{2\mu}-\frac{e^2}{|q^{\tt H}|}  \;,
\eeqa
and $K^{\tt H}_i$ is the Runge-Lenz vector
\beqa  K^{\tt H}_i&=& \frac{1}{\mu e^2} \epsilon_{ijk}  L^{\tt H}_jp^{\tt H}_k  +\frac{q^{\tt H}_i}{|q^{\tt H}|}
\eeqa
$q^{\tt H}_i$ and $p^{\tt H}_i$ are canonically conjugate phase space  variables $\{q^{\tt H}_i, p^{\tt H}_j\} =\delta_{ij}$,  the angular momentum is as usual given by $L^{\tt H}_i=\epsilon_{ijk} q^{\tt H}_j p^{\tt H}_k$ and we restrict to bound states $H<0$.
$\mu$ and $e$ are, respectively, the reduced mass and charge.
The Poisson brackets for  $L^{\tt H}_i$ and
   $x^{\tt H}_i$ are given by 
\beqa  \{ x^{\tt H}_i, x^{\tt H}_j\} &=&\frac 2{\mu e^4}\epsilon_{ijk} L^{\tt H}_k\cr & &\cr\{ x^{\tt H}_i, L^{\tt H}_j\} &=&\epsilon_{ijk} x^{\tt H}_k\cr & &\cr \{ L^{\tt H}_i, L^{\tt H}_j\} &=&\epsilon_{ijk} L^{\tt H}_k\label{hatmalg}\eeqa 
In comparing with (\ref{snydrsubalg}) and (\ref{angmomalg}), we see that here  $\sqrt{\frac \mu 2}\; e^2$ plays the role of the deformation parameter $\Lambda$ in the Snyder algebra. Also, like in (\ref{xdotl}),  the scalar product of $L^{\tt H}_i$ and $x^{\tt H}_i$ vanishes, and so just as with the Snyder algebra, there is only one independent  quadratic invariant of $SO(4)$.\footnote{We thank Tom Kephart for pointing this out.}  The quadratic Casimir is
$\Bigl(L^{\tt H}_i\pm \sqrt{\frac \mu 2}\; e^2x^{\tt H}_i\Bigr)^2 =L^{\tt H}_iL^{\tt H}_i+\frac {\mu e^4}2x^{\tt H}_ix^{\tt H}_i$, and its eigenvalues  $\hbar^2 j(j+1)$  label the irreducible representations of the $SO(4)$ algebra occurring in the quantum theory. The index $j$  is related to  the principal quantum number $n_p$ according to $n_p=2j+1$. \cite{Bander:1965rz}

In the Snyder algebra, the action of the $SO(4)$ generators on momentum  is given by the commutation relations  (\ref{snydrsubalg}) and (\ref{angmomalg}).  One gets  a more complicated  action of $SO(4)$  on momentum in the   hydrogen atom system.  The  Poisson brackets of the generators  with the momenta lead to a mixing with the position variables $q^{\tt H}_i$,
\beqa   \{ x^{\tt H}_i, p^{\tt H}_j\}&=& \frac 1{\sqrt{-H}}\;\biggl(\frac{p^{\tt H}_ip^{\tt H}_j-\delta_{ij}|p^{\tt H}|^2}{\mu e^2} -
\frac{q^{\tt H}_iq^{\tt H}_j- \delta_{ij}|q^{\tt H}|^2}{|q^{\tt H}|^3}+ \frac{e^2  x^{\tt H}_i q^{\tt H}_j }{2 \sqrt{-H} \;|q^{\tt H}|^3}\biggr)\cr & &\cr \{ L^{\tt H}_i,p^{\tt H} _j\}&=&\epsilon_{ijk}p^{\tt H}_k \eeqa
 The energy  eigenfunctions for the hydrogen atom in momentum space were found long ago by  Podolsky and Pauling\cite{Pauling} and by Fock\cite{Fock:1935vv}.  They are spherical harmonics on $S^3$, and so are  given by the expressions (\ref{eignfnphi}) [with $\alpha=0$].
The three-sphere is not obtained in this case using the map (\ref{coordnitize}).  Rather, it is  obtained by an inverse stereographic projection of momentum space.  Here the four-dimensional embedding coordinates $P_A,\; A=1,2,3,4$, $\;P_AP_A=1$, associated with any given bound state energy $E_n$ are  given by\cite{Bander:1965rz}
\be P_i =\frac{2p^{\tt H}_i\sqrt{-2mE_n}}{{|p^{\tt H}}|^2-2mE_n}   \qquad\quad P_4 =\frac{{|p^{\tt H}}|^2+2mE_n} {{|p^{\tt H}}|^2-2mE_n}\ee
Unlike (\ref{coordnitize}), this map takes ${\mathbb{R}}^{3}$ to all of $S^3$.  Setting $ P_4 =\cos\chi$ as before, $\chi$ now ranges from $0 $ to $\pi$, rather than  $0 $ to $\pi/2$, as with  the eigenfunctions  (\ref{eignfnphi}) for 
 the Snyder algebra.
  We argued in section 4 that there were two distinct sets of orthonormal eigenfunctions for the Snyder algebra, and that  they correspond to $j$   either all integers or all half-integers.
 Here, however, since the domain is $0 \le \chi\le\pi$ for the hydrogen atom system, the Gegenbauer polynomials 
 $\{C^{(\ell+1)}_{n}(\cos\chi),\;n=0,1,2,...\}$ forms one complete orthonormal set, and using $n=2j-\ell$, 
  the spherical harmonics  with both integer and half-integer values for $j$ span the Hilbert space.  Unlike in the subsection  4.1, the basis functions now consist of all irreducible matrix representations of $SU(2)$,  $\{{\tt D}^{j}(g),\; j=0,\frac 12,1,\frac 32,...\}$, which  allows  the principal quantum number $n_p=2j+1$ to be any positive integer.


\bigskip


 \begin{thebibliography}{99}

\bibitem{Snyder:1946qz}
  H.~S.~Snyder,
 ``Quantized space-time,''
  Phys.\ Rev.\  {\bf 71}, 38 (1947).

\bibitem{'tHooft:1996uc}
  G.~'t Hooft,
  ``Quantization of point particles in (2+1)-dimensional gravity and space-time
  discreteness,''
  Class.\ Quant.\ Grav.\  {\bf 13}, 1023 (1996).


\bibitem{Matschull:1997du}
  H.~J.~Matschull and M.~Welling,
  ``Quantum mechanics of a point particle in (2+1)-dimensional gravity,''
  Class.\ Quant.\ Grav.\  {\bf 15}, 2981 (1998).


\bibitem{leilu} Lei Lu and A. Stern, ``Particle dynamics on Snyder space,"  in preparation.

\bibitem{Mignemi:2011gr}
  S.~Mignemi,
  ``Classical and quantum mechanics of the nonrelativistic Snyder model,''
  Phys.\ Rev.\  {\bf D84}, 025021 (2011).


\bibitem{Bander:1965rz}  For  the case of the hydrogen atom, see 
  M.~Bander and C.~Itzykson,
  ``Group theory and the hydrogen atom,''
  Rev.\ Mod.\ Phys.\  {\bf 38}, 330 (1966).

\bibitem{Pauling}
B.~ Podolsky and L.~ Pauling, ``The Momentum Distribution in Hydrogen-Like Atoms,'' Phys.\ Rev.\ {\bf 34}, 109 (1929). 

 \bibitem{Fock:1935vv}
  V.~Fock,
  ``On the Theory of the hydrogen atoms,''
  Z.\ Phys.\  {\bf 98}, 145 (1935).

\bibitem{AmelinoCamelia:2010pd}
 For a recent review see,  G.~Amelino-Camelia,
  ``Doubly-Special Relativity: Facts, Myths and Some Key Open Issues,''
  Symmetry {\bf 2}, 230 (2010).

 

\bibitem{Romero:2008jf}
  J.~M.~Romero and A.~Zamora,
  ``The Area Quantum and Snyder Space,''
  Phys.\ Lett.\  B {\bf 661}, 11 (2008).
 
\bibitem{Stern:2010ri}
  A.~Stern,
  ``Relativistic Particles on Quantum Space-time,''  Phys.\ Lett.\  A {\bf 375}, 2498 (2011).

\bibitem{Hellund:1954zz}
  E.~J.~Hellund and K.~Tanaka,
  ``Quantized Space-Time,''
  Phys.\ Rev.\  {\bf 94}, 192 (1954).




\bibitem{Balachandran:1991zj} See for example,
  A.~P.~Balachandran, G.~Marmo, B.~S.~Skagerstam and A.~Stern,
  ``Classical topology and quantum states,''
{\it  Singapore, Singapore: World Scientific (1991) 358 p}.

 



\bibitem{tolman} 
J.D. Talman, `Special functions : a group theoretic approach',  Benjamin, New York, 1968.


\bibitem{Battisti:2010sr}
  M.~V.~Battisti and S.~Meljanac,
  ``Scalar Field Theory on Non-commutative Snyder Space-Time,''
  Phys.\ Rev.\  D {\bf 82}, 024028 (2010).

\bibitem{Girelli:2010wi}
  F.~Girelli and E.~R.~Livine,
  ``Scalar field theory in Snyder space-time: Alternatives,''
  JHEP {\bf 1103}, 132 (2011).

  \bibitem{Jaroszkiewicz} G. Jaroszkiewicz, ``A dynamical model for the origin of Snyder's quantized spacetime algebra,''   J. Phys. A: Math. Gen. {\bf 28} L343 (1995).

\bibitem{Romero:2004er}
  J.~M.~Romero and A.~Zamora,
  ``Snyder noncommutative space-time from two-time physics,''
  Phys.\ Rev.\  D {\bf 70}, 105006 (2004).


\bibitem{Romero:2006pe}
  J.~M.~Romero and J.~D.~Vergara,
  ``The Parametrized relativistic particle and the Snyder space-time,''
  arXiv:hep-th/0602058.


\bibitem{Banerjee:2006wf}
  R.~Banerjee, S.~Kulkarni and S.~Samanta,
  ``Deformed symmetry in Snyder space and relativistic particle dynamics,''
  JHEP {\bf 0605}, 077 (2006).



\bibitem{Chatterjee:2008bp}
  C.~Chatterjee and S.~Gangopadhyay,
  ``Kappa-Minkowski and Snyder algebra from reparametrisation symmetry,''
  Europhys.\ Lett.\  {\bf 83}, 21002 (2008).


 
  \end{thebibliography}
\end{document}